\documentclass[runningheads]{llncs}
\usepackage[T1]{fontenc}
\usepackage{graphicx}
\begin{document}
\title{The Persistence of Contrarianism on Twitter: \\ Mapping users' sharing habits for the Ukraine war, COVID-19 vaccination, and the 2022 Midterm Elections}
\titlerunning{Persistence of contrarianism on Twitter}
%
\author{David Axelrod \orcidID{0000-0002-9840-8711} (\email{daaxelro@iu.edu}) \and \\
Sangyeon Kim\ \orcidID{0000-0001-9105-8598} (\email{ski15@iu.edu}) \and \\
John Paolillo\ \orcidID{0009-0000-0876-5778} (\email{paolillo@indiana.edu})}
\authorrunning{D. Axelrod, et al.}

\institute{Indiana University Observatory on Social Media}

\maketitle              
\begin{abstract}
Empirical studies of online disinformation emphasize matters of public concern such as the COVID-19 pandemic, foreign election interference, and the Russo-Ukraine war, largely in studies that treat the topics separately. Comparatively fewer studies attempt to relate such disparate topics and address the extent to which they share behaviors. In this study, we compare three samples of Twitter data on COVID-19 vaccination, the Ukraine war and the 2022 midterm elections, to ascertain how distinct ideological stances of users across the three samples might be related. Our results indicate the emergence of a broad contrarian stance that is defined by its opposition to public health narratives/policies along with the Biden administration's foreign policy stances. Sharing activity within the contrarian position falls on a spectrum with outright conspiratorial content on one end. We confirm the existence of ideologically coherent cross-subject stances among Twitter users, but in a manner not squarely aligned with right-left political orientations. 
\end{abstract}
\section{Introduction}
Overwhelmingly, data collection efforts \cite{deverna2021covaxxy,aiyappa2023multi} and empirical studies of online disinformation tend to address a single topic at a time, such as the COVID-19 pandemic \cite{cov_twitter_v_facebook}, foreign election interference \cite{tuckerRussia}, and the Russo-Ukraine war \cite{pierriUkr}. In addition, disciplines tend to emphasize narrow topical foci, e.g., in the case of acceptance of COVID-19 vaccines and public health guidance, even when their ideological resonance is clear \cite{misinfo_vacc}. When ideology is brought into focus, studies tend to rely on text-based methods \cite{rao2010classifying,preoctiuc2017beyond,stefanov2020predicting,darwish2020unsupervised}, whereas a network-focused study might better identify shared attitudes across topics \cite{barbera2015birds,conover2011predicting,gu2016ideology,wong2016quantifying}. We build on this work by asking: how might users' ideological stances on these three topics be related? To the extent that they correlate, is this reflective of overall political orientation? And how is the potential for misinformation distributed across the stances? 

In this study, we address these questions by comparing three samples of Twitter data on COVID-19 vaccination, the Ukraine war and the 2022 midterm elections. COVID-19 and Ukraine war are two of the largest global events of the last five years well represented on Twitter; the US 2022 midterm elections took place during the pandemic, and its inclusion alongside the other two helps situate them politically with respect to the US context. We adopt an approach employing Principal Components and cluster analysis to analyze the network of retweets, focusing in particular on a subset of users common to all three samples. 
\section{Data}
This analysis employs two published Twitter datasets about COVID-19 vaccines \cite{deverna2021covaxxy} and the 2022 United States midterm elections \cite{aiyappa2023multi} along with a third dataset on the Ukraine war that was collected using a similar method. For all three samples, data collection begins using a limited set of topic-relevant terms e.g., ``Ukraine''; the initial collection is examined for additional relevant terms to expand the search query. The resulting datasets span January 4, 2021 to February 7, 2023 (for COVID-19), January 28, 2022 to December 2, 2022 (Ukraine), and September 10, 2022 to December 31, 2022 (midterm). Because these datasets contain varying amounts of spam and unrelated content, we further trimmed the dataset to retweets of persistently active users, identified by checking against a list of users active during October 2023. This results in retweets by 930,277 users in the COVID-19 data, 1,532,489 users in the Ukraine data, and 466,925 users in the midterm data. 
\section{Methods}
For each of the constituent datasets, we performed an analysis on retweets of influencers, operationalized as users who received retweets from at least 0.1\% of the user set (retweeters). Checks were done for varying thresholds, including none; the 0.1\% threshold results in a considerable reduction in noise. Subsequent analyses identifying structure in users' retweeting patterns was accomplished with Principal Components Analysis (PCA) followed by cluster analysis.
\subsection{Principal Components Analysis}\label{pca}
Briefly, PCA is a statistical procedure for reducing a high-dimension dataset to a smaller number of latent dimensions of shared variation \cite{basilevsky2009statistical}. Here, we apply it to the matrix of retweets, in which influencers are represented as columns (i.e., variables) and retweeting users are rows (observations); each cell is scored 1 or zero for whether or not the user retweeted the influencer. 
PCs are linear sums of the original variables; this allows us to take PCAs conducted independently, e.g., for different samples, and combine them by conducting a PCA across the scores for matching retweeters. In the present analysis, we do this in three steps. First, to make analysis more tractable, we split the samples into 1-week rolling windows for PCA. The window PCA scores are then combined into a matrix on which the sample PCA is conducted.
Lastly, a PCA is done across samples: users participating in all the samples are matched, their scores extracted and combined into a matrix on which we conduct the final PCA, which we refer to as the ``common space''. Henceforth, when we refer to PCs, these are from the common space, unless explicitly identified as belonging to a sample. Examining the sample PC loadings on the common space PCs allows us to identify relationships among the samples. Ideological interpretations are developed by referring to the most-often retweeted influencers across the three samples. Out of 2,200,314 retweeters, 184,627 are shared across all three samples. 

For dimension reduction, some number of PCs less than the number of original variables is used; the variation belonging to the remaining dimensions are treated as ``error''. In the window PCAs, we retain up to a maximum of 10 PCs to avoid discarding within-sample variation that might correlate across samples. This yielded 3,810 window-PCs for the COVID-19 sample, 1,540 for the Ukraine sample, and 1,110 for the midterm sample. In the sample PCAs, PCs were extracted to cover 95\% of the variation in the respective matrix, yielding 808 PCs for the COVID-19 sample, 327 PCs for Ukraine, and 203 PCs for Midterm.  In the common space PCA, six PCs were extracted by a scree test, covering 46.7\% of the variation. These were submitted for cluster analysis and interpretation.
\subsection{Clustering}
Since the activity levels of users in the COVID, Ukraine, and midterm samples is highly skewed, with most users being relatively inactive within and especially across the samples, we focus our analysis on the more active users present in all three samples. We accomplish this by measuring the Euclidean distance from the common-space origin for users present in all three sets, retaining 17,366 users at the $90^{th}$ distance percentile or greater, hence removing users located near the origin, whose cosine values could exaggerate their similarity to users placed further away. The user-user cosine distance matrix was passed to HDBSCAN\cite{mcinnes2017accelerated}, a density based clustering algorithm, with minimum cluster size set to 20 users. This solution produced the minimal number of clusters (23) compared to other tested approaches, with six containing 500 users or more; the remaining clusters reflect more idiosyncratic post-sharing behaviors. For subsequent interpretation, we focus on these six high-membership clusters, henceforth referred to as A (red, 3,058 members), B (blue, 818), C (orange, 3,103), D (brown, 508), E (green, 2,765), and F (violet, 503).
\begin{figure}
\centering
\includegraphics[width=0.8\textwidth]{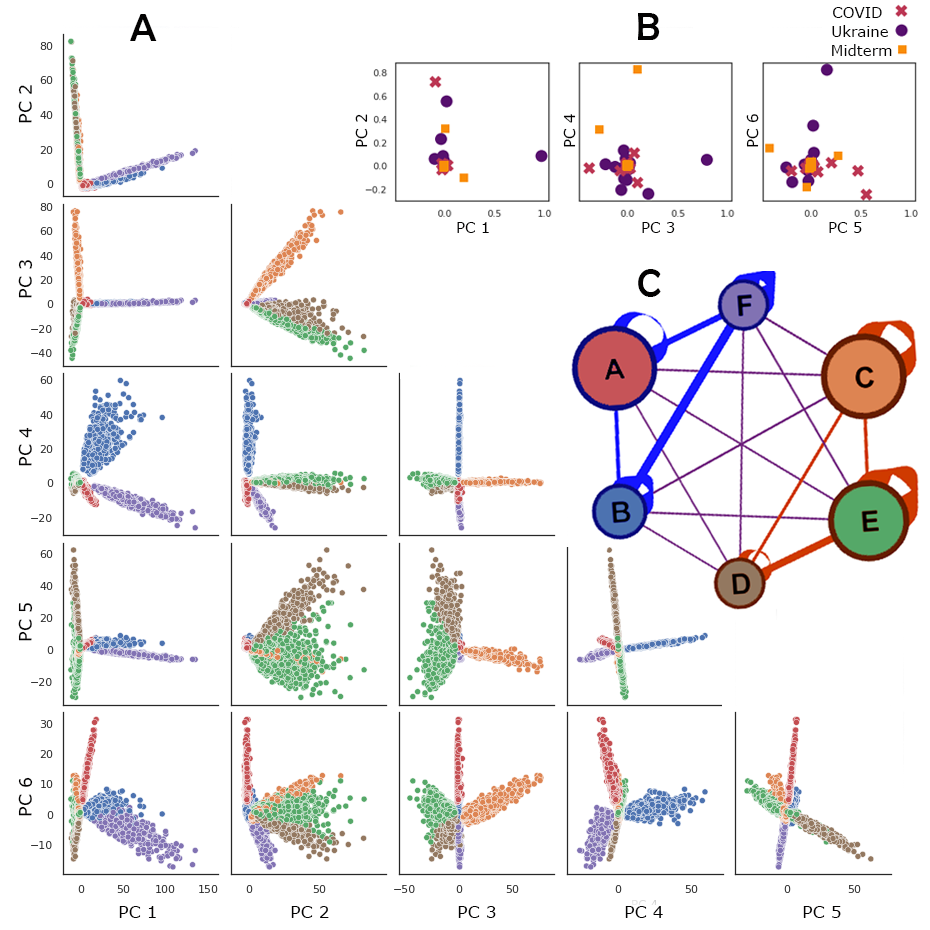}
\caption{Six largest user groups identified by clustering on principal component scores from the cross-sample common space (see section \ref{pca}). Clusters are color encoded equivalently in 1(A) and 1(C). (A) Pair plot of clustered users along the common space principal components. (B) Biplots showing the rotations of the sample-derived principal components into the common space. (C) Co-retweet network for the six main clusters with edge weights proportional to the number of shared retweets and edge colors highlighting the two-chamber structure pulled out by PC1/PC2 and Louvain community detection.} \label{fig1}
\end{figure}
\section{Results}
The first two PCs separate the Biden administration-aligned clusters A, B and F from the contrarian clusters C, D and E (see Figure 1(A)). The first group is spread out along PC1, which is overwhelmingly defined by the loading of Ukraine sample PC1 (Figure 1(B), first panel). Clusters B and F are opposed on PC4, for which midterm PC1 has the highest loading (Figure 1(B) panel 2). Cluster B emphasizes domestic anti-Trump content (POTUS [the Biden administration], Mary L Trump, Rep. Eric Swalwell, Brian Tyler Cohen, Robert Reich, Amy Klobuchar, etc.) whereas F emphasizes international media (e.g., BBC, Kyiv Independent, Garry Kasparov, Mariya Dekhtyaruk, Auschwitz Memorial, etc.). Cluster A only fully separates from B and F on PC6 (Ukraine PC3), which is notable given that cluster A's members retweet primarily institutional US left media (e.g., NBC News, NY Times, Washington Post, the Daily Beast, Mehdi Hassan, etc.) As indicated by the network in Figure 1(C), the three clusters share some retweeting behavior especially B and F, but have different foci of attention.

Similarly, clusters C, D and E assemble tightly together on PC2 (COVID PC1 and Ukraine PC2) and form a distinct retweet-sharing sub-network. These clusters represent contrarian perspectives unified in their opposition to COVID public health policies, which stands out given the the absence of a pro-vaccine defined cluster among the administration-aligned clusters. Although D and E is dominated by retweets of influencers on the US political right, cluster C favors left-oriented anti-US/anti-imperialist perspectives (Glenn Greenwald, Max Blumenthal, Kim Dotcom, etc.). Cluster C separates from D and E on PC3 (Ukraine PC2), and is most nearly opposite cluster D. Clusters D and E are opposed on PC5, (COVID PC1, PC2 and midterm PC1); D favors personalities whose focus was opposition to COVID policies (Dr. Aseem Malhotra, Robert Kennedy Jr., Brook Jackson, etc.) whereas E covers a broader libertarian/right political framing (Elon Musk, Kari Lake, Phillip Buchanan [catturd2], Charlie Kirk, etc.). Cluster D is also opposite Cluster A on PC6 (Ukraine PC3), with the former featuring Paul Joseph Watson (prisonplanet) and RadioGenova, an account focused on opposing immigration to Europe. On the other hand, cluster A relies heavily on an Oxford University professor specializing in international relations.

\section{Conclusion}
We observe two main stances representing contrarian and institutional orientations that align with the US political left and right 
 in five out of the six main clusters. These stances are durable and correlate strongly across the three samples. Within the contrarian stance, cluster C is notably split from D and E earliest of any of the sub-divisions, suggesting that among otherwise contrarian-minded users, they tend to be disjoint in their selection of sources, e.g., an anti-imperialist/anti-United States framing may not be palatable to users on the political right. International viewpoints are largely pushed to the margins in our samples, likely due to the dominance of English on Twitter and the United States within the Anglosphere. Notably, while there is a clear pro-Ukraine cluster (and PC), we do not find a distinct, specifically pro-vaccine cluster.

While this work attempts to triangulate the ideological space for Twitter users across three pivotal topics, our conclusions apply to a limited subset of particularly motivated users. Most users in our data were less active than those we focus on in our clustering and their ideological distributions may differ. Nonetheless, focusing on motivated users who participate across multiple samples is informative for understanding how active users wield a capacity to dominate the discussion around important topics.
\begin{credits}
\subsubsection{\ackname} This study was supported in part by a grant from the Knight Foundation to the Indiana University Observatory on Social Media (OSoMe).
\end{credits}
%
%
%
 \bibliographystyle{splncs04}
 \bibliography{version1}

\begin{thebibliography}{10}
\providecommand{\url}[1]{\texttt{#1}}
\providecommand{\urlprefix}{URL }
\providecommand{\doi}[1]{https://doi.org/#1}

\bibitem{aiyappa2023multi}
Aiyappa, R., DeVerna, M.R., Pote, M., Truong, B.T., Zhao, W., Axelrod, D., Pessianzadeh, A., Kachwala, Z., Kim, M., Seckin, O.C., et~al.: A multi-platform collection of social media posts about the 2022 us midterm elections. In: Proceedings of the international AAAI conference on web and social media. vol.~17, pp. 981--989 (2023)

\bibitem{barbera2015birds}
Barber{\'a}, P.: Birds of the same feather tweet together: Bayesian ideal point estimation using twitter data. Political analysis  \textbf{23}(1),  76--91 (2015)

\bibitem{basilevsky2009statistical}
Basilevsky, A.T.: Statistical factor analysis and related methods: theory and applications. John Wiley \& Sons (2009)

\bibitem{conover2011predicting}
Conover, M.D., Gon{\c{c}}alves, B., Ratkiewicz, J., Flammini, A., Menczer, F.: Predicting the political alignment of twitter users. In: 2011 IEEE third international conference on privacy, security, risk and trust and 2011 IEEE third international conference on social computing. pp. 192--199. IEEE (2011)

\bibitem{darwish2020unsupervised}
Darwish, K., Stefanov, P., Aupetit, M., Nakov, P.: Unsupervised user stance detection on twitter. In: Proceedings of the international AAAI conference on web and social media. vol.~14, pp. 141--152 (2020)

\bibitem{deverna2021covaxxy}
DeVerna, M.R., Pierri, F., Truong, B.T., Bollenbacher, J., Axelrod, D., Loynes, N., Torres-Lugo, C., Yang, K.C., Menczer, F., Bryden, J.: Covaxxy: A collection of english-language twitter posts about covid-19 vaccines. In: Proceedings of the international AAAI conference on web and social media. vol.~15, pp. 992--999 (2021)

\bibitem{tuckerRussia}
Eady, G., Paskhalis, T., Zilinsky, J., Bonneau, R., Nagler, J., Tucker, J.: Exposure to the russian internet research agency foreign influence campaign on twitter in the 2016 us election and its relationship to attitudes and voting behavior. Nature Communications  \textbf{14}(62) (2023)

\bibitem{gu2016ideology}
Gu, Y., Chen, T., Sun, Y., Wang, B.: Ideology detection for twitter users with heterogeneous types of links. arXiv preprint arXiv:1612.08207  (2016)

\bibitem{mcinnes2017accelerated}
McInnes, L., Healy, J.: Accelerated hierarchical density based clustering. In: Data Mining Workshops (ICDMW), 2017 IEEE International Conference on. pp. 33--42. IEEE (2017)

\bibitem{pierriUkr}
Pierri, F., Luceri, L., Jindal, N., Ferrara, E.: Propaganda and misinformation on facebook and twitter during the russian invasion of ukraine. In: Proceedings of the 15th ACM Web Science Conference 2023. p. 65–74. WebSci '23, Association for Computing Machinery, New York, NY, USA (2023), \url{https://doi.org/10.1145/3578503.3583597}

\bibitem{misinfo_vacc}
Pierri, F., Perry, B., DeVerna, M., Yang, K.C., Flammini, A., Menczer, F., Bryden, J.: Online misinformation is linked to early covid-19 vaccination hesitancy and refusal. Scientific Reports  \textbf{12}(5966) (2022). \doi{10.1038/s41598-022-10070}, \url{https://doi.org/10.1038/s41598-022-10070-w}

\bibitem{preoctiuc2017beyond}
Preo{\c{t}}iuc-Pietro, D., Liu, Y., Hopkins, D., Ungar, L.: Beyond binary labels: Political ideology prediction of twitter users. In: Proceedings of the 55th annual meeting of the association for computational linguistics (volume 1: long papers). pp. 729--740 (2017)

\bibitem{rao2010classifying}
Rao, D., Yarowsky, D., Shreevats, A., Gupta, M.: Classifying latent user attributes in twitter. In: Proceedings of the 2nd international workshop on Search and mining user-generated contents. pp. 37--44 (2010)

\bibitem{stefanov2020predicting}
Stefanov, P., Darwish, K., Atanasov, A., Nakov, P.: Predicting the topical stance and political leaning of media using tweets. In: Proceedings of the 58th Annual Meeting of the Association for Computational Linguistics. pp. 527--537 (2020)

\bibitem{wong2016quantifying}
Wong, F.M.F., Tan, C.W., Sen, S., Chiang, M.: Quantifying political leaning from tweets, retweets, and retweeters. IEEE transactions on knowledge and data engineering  \textbf{28}(8),  2158--2172 (2016)

\bibitem{cov_twitter_v_facebook}
Yang, K.C., Pierri, F., Hui, P.M., Axelrod, D., Torres-Lugo, C., Bryden, J., Menczer, F.: The covid-19 infodemic: Twitter versus facebook. Big Data \& Society  \textbf{8}(1),  20539517211013861 (2021). \doi{10.1177/20539517211013861}, \url{https://doi.org/10.1177/20539517211013861}

\end{thebibliography}
\end{document}